\documentclass[aps,prb,amsmath,amssymb,twocolumn,showpacs,superscriptaddress]{revtex4-1}
\usepackage[pdftex]{graphicx}
\usepackage[colorlinks=true,citecolor=blue]{hyperref}
\usepackage{braket}

\newcommand{\tx}{\text}

\renewcommand{\vec}[1]{\ensuremath{\mathbf{#1}}}

\newcommand{\Gm}{\Gamma}

\newcommand{\dbraket}[1]{\ensuremath{ \langle \! \langle #1 \rangle \! \rangle}}

\begin{document}
\title{Short-time counting statistics of charge transfer in Coulomb-blockade systems}

\author{Philipp Stegmann} \email{philipp.stegmann@uni-due.de}
\affiliation{Theoretische Physik and CENIDE, Universit\"at Duisburg-Essen, D-47048 Duisburg, Germany}

\author{J\"urgen K\"onig}
\affiliation{Theoretische Physik and CENIDE, Universit\"at Duisburg-Essen, D-47048 Duisburg, Germany}\date{\today}

\begin{abstract}
We study full counting statistics of electron tunneling in Coulomb-blockade systems in the limit of short measuring-time intervals.
This limit is particularly suited to identify correlations among tunneling events, but only when analyzing the charge-transfer statistics in terms of factorial cumulants $C_{{\rm F}, m}(t)$ rather than ordinary ones commonly used in literature.
In the absence of correlations, the short-time behavior of the factorial cumulants is given by $C_{{\rm F}, m}(t) \propto (-1)^{m-1}t^m$.
A different sign and/or a different power law of the time dependence indicates correlations.
We illustrate this for sequential and Andreev tunneling in a metallic single-electron box.
\end{abstract}

\maketitle

% - - - - - - - - - - - - - - - - - - - - - - - - - - - - - - - - - - - - - -
\section{Introduction}
The transfer of electrons through a tunneling barrier is a stochastic process.
If the individual tunneling events are {\it uncorrelated} (i.e., {\it independent} of each other) and occur with the {\it same} single-particle tunneling probability $p(t)=\Gamma t$, the probability $P_N(t)$ that $N\ge0 $ electrons have been transferred during the time interval $[0,t]$ is described by the Poisson distribution, $P_N(t)= e^{-p} p^N/N!$.
The more general scenario of {\it uncorrelated} ({\it independent}) but {\it nonidentical} tunneling events is characterized by the Poisson binomial distribution\cite{wang_number_1993}
\begin{equation}
\label{eq:poissonbinomialdist}
	P_N(t) = \sum_{A_N} \prod_{j \in A_N}p_j(t)  \prod_{k \in A_N^\tx{c}} \left[1-p_k(t) \right] \, ,
\end{equation}
where individual tunneling trials $j$ succeed with probability $p_j(t)$ or fail with $1-p_j(t)$. 
The sum goes over all subsets $A_N$ of the natural numbers with cardinality $N$, and $A_N^\tx{c}$ is the complement of $A_N$.
While a single number $p$ is sufficient to fix the Poisson distribution, the huge number of possible sets of $\{ p_j \}$ with the restriction $0\le p_j\le1$ generates such a plethora of possible charge-transfer distributions $P_N(t)$, that one may wonder whether it is possible to approximate {\it any} tunneling statistics $P_N(t)$ by a Poisson binomial distribution with properly chosen $p_j$'s.
This is, however, not the case: if the tunneling events are {\it correlated}, i.e., they do {\it not} occur independently from each other, then the charge-transfer statistics can be qualitatively changed.

In this paper, we demonstrate that the short-time limit is particularly suited to identify correlations, but only when studying {\it factorial} cumulants of $P_N(t)$ rather than ordinary ones commonly used in literature.
For this, we consider Coulomb-blockade systems, in which the electrostatic potential of a confined region such as a quantum dot or a metallic island depends on the number of contained electrons.
The latter can be detected by a sensitive electrometer, e.g., an electrostatically coupled quantum point contact~\cite{gustavsson_counting_2005, fujisawa_bidirectional_2006, gustavsson_measurements_2007, fricke_bimodal_2007, flindt_universal_2009, gustavsson_electron_2009, fricke_high_2010, fricke_high-order_2010, komijani_counting_2013} or single-electron transistor,\cite{martinis_metrological_1994, dresselhaus_measurement_1994, lotkhov_storage_1999, lu_real_2003, bylander_current_2005} but also optical,~\cite{kurzmann_optical_2016} interferometric~\cite{dasenbrook_dynamical_2016} or other detection schemes are conceivable. 
Changes in the electron number indicate the individual tunneling events, from which one can calculate the full counting statistics $P_N(t)$.~\cite{levitov_electron_1996, bagrets_full_2003}
The latter provides information about the impact of ferromagnetic leads,\cite{lindebaum_spin-induced_2009} electron-phonon interactions,\cite{schmidt_charge_2009, souto_transient_2005} non-Markovian,\cite{braggio_full_2006, flindt_counting_2008, flindt_counting_2010} and frequency-dependent\cite{emary_frequency-dependent_2007, ubbelohde_measurement_2012} effects. 
Various Coulomb-blockade systems including multilevel quantum dots,\cite{belzig_full_2005} interferometers,\cite{urban_coulomb-interaction_2008} multistable,\cite{schaller_counting_2010} and feedback-controlled systems\cite{poeltl_feedback_2011, daryanoosh_stochastic_2016, wagner_squeezing_2016} have been investigated.

The Coulomb interaction among the electrons on the island can introduce correlations since the probability of a given tunneling event may depend on the island's charge state. Moreover, superconducting pairing interaction\cite{belzig_full_counting_2001, boerlin_full_2002, cuevas_full_2003, johansson_full_2003, pilgram_noise_2005, morten_full_2008, braggio_superconducting_2011, soller_full_2014} becomes relevant if the island is coupled to a superconducting lead.
Then, two electrons can be transferred simultaneously between island and Cooper-pair condensate of the lead, a process referred to as Andreev tunneling.~\cite{andreev_thermal_1964}

The outline of the paper is as follows.
First, in Sec.~\ref{sec:ordinarystatistics}, we introduce ordinary and factorial cumulants. Then, in Sec.~\ref{sec:factorialstatics}, we study their properties in the short-time limit. We explain why ordinary cumulants are inappropriate to detect the presence of correlations in the short-time limit, in contrast to factorial and generalized factorial cumulants.
Finally, we illustrate this claim in Sec.~\ref{sec:model} for a model system, a single-electron box consisting of a normal-metal island tunnel coupled to a superconductor, for which counting statistics of Andreev tunneling has recently been measured.~\cite{saira_environmentally_2010, maisi_real_2011, maisi_full_2014}

% - - - - - - - - - - - - - - - - - - - - - - - - - - - - - - - - - - - - - -
\section{Moments and cumulants}\label{sec:ordinarystatistics}
Full counting statistics is commonly characterized by ordinary moments $\langle N^m\rangle (t):= \sum_N N^m P_N(t)$ or ordinary cumulants $C_m(t):= \dbraket {N^m}  (t)$.
They are obtained from the generating function ${\cal M}(z,t) := \sum_N e^{N z} P_N(t)$ via performing the derivatives $\langle N^m\rangle (t) =\partial_z^m {\cal M}(z,t)|_{z=0}$ and $\dbraket {N^m} (t) := \partial_z^m [{\ln \cal M}(z,t)] |_{z=0}$.
Due to the logarithm, cumulants (unlike moments) of independent charge-transfer channels simply add up.

Alternatively, full counting statistics can be characterized by factorial moments and cumulants. Factorial moments $\langle N^{(m)} \rangle (t) = \sum_N N^{(m)} P_N(t)$ are defined as the expectation values of {\it factorial powers} $N^{(m)}:= N(N-1)\cdots (N-m+1)$ and can be derived via $\langle N^{(m)} \rangle (t) = \partial_z^m {\cal M}_{\rm F}(z,t)|_{z=0}$ from the corresponding generating function,
\begin{equation}
\label{eq:generating_factorial}
	{\cal M}_{\rm F}(z,t) := \sum_N(z+1)^N P_N(t) \, .
\end{equation}
Per definition, we do the counting in such a way that the number $N$ is always $\ge 0$: we count only the electrons that {\it leave} the central part of the system (e.g. quantum dot or island) into some leads and not those entering (i.e., $N$ is in general \emph{not} the {\it net} charge transfer between island and lead).

Factorial cumulants are obtained from $C_{{\rm F}, m}(t):= \dbraket {N^{(m)} } (t) := \partial_z^m [\ln {\cal M}_{\rm F}(z,t)] |_{z=0}$. 
As compared to the generating function for ordinary moments and cumulants, $e^{Nz}$ is replaced by $(z+1)^N$.
Let us denote by $z_j(t)$ the zeros of the generating function, ${\cal M}_{\rm F}(z_j(t),t)=0$.
Together with the normalization ${\cal M}_{\rm F}(0,t)=1$, this yields ${\cal M}_{\rm F}(z,t) = \prod_j \left[ 1-z/z_j(t) \right]$ (multiple zeros appear with corresponding multiplicity in the product).
If we now define $p_j(t) := - 1/z_j(t)$, then\cite{remark0}
\begin{equation}\label{eq:prodzj}
	{\cal M}_{\rm F}(z,t) = \prod_j \left[ 1-p_j(t)+(z+1)p_j(t) \right]\, .
\end{equation}
Thus, the charge-transfer statistics $P_N(t)$ acquires the form of Eq.~(\ref{eq:poissonbinomialdist}), and
the factorial cumulants are given by
\begin{equation}
\label{eq:cumwithpj}
	C_{{\rm F}, m}(t)=(-1)^{m-1}(m-1)! \sum_j \left[ p_j(t) \right]^m \, .
\end{equation}
It is important, however, to emphasize that a mapping onto a Poisson binomial distribution for independent electron tunneling events is only possible if all the $p_j(t)$'s are real and lie between $0$ and $1$. Then, the sign of $C_{{\rm F}, m}(t)$ is fixed, $(-1)^{m-1}C_{{\rm F}, m}(t)\ge 0$.
In general, correlations influence the zeros of the generating function such that the $p_j(t)$'s can be $<0$ or $>1$ and pairs of complex-valued $p_j(t)$'s can appear.
This can change the sign of the factorial cumulants.\cite{kambly_factorial_2011, kambly_time-dependent_2013, stegmann_detection_2015, droste_2016}
Recently, we have shown\cite{stegmann_detection_2015} that the detection of the presence of correlations can be tremendously facilitated via 
shifting the complex variable~$z$ by the amount $s-1$ along the real axis in the complex plane.
This yields
\begin{equation}
	{\cal M}_s(z,t):=\frac{\sum\limits_{N=0}^\infty (z+s)^N P_N(t)}{\sum\limits_{N=0}^\infty s^N P_N(t)} \, ,
\end{equation}
where the denominator guarantees the normalization ${\cal M}_s(0,t) = 1$.
The factorial generating function ${\cal M}_{\rm F}(z,t)$ is contained as the special case $s=1$.
The generalized factorial cumulants take the form
\begin{equation}
\label{eq:cumulantscos}
	C_{s, m}(t) = (-1)^{m-1} (m-1)! \sum_j \left[- \frac{1}{z_j(t)-s+1} \right]^m
\end{equation}
and a violation of $(-1)^{m-1}C_{s, m}(t) \ge 0$
for arbitrary orders $m$ if $s\ge0$ and even orders $m$ if $s<0$ indicates correlations.~\cite{stegmann_detection_2015}

We emphasize that the derivations in this section are valid for arbitrary distributions $P_N(t)$, independent of the underlying physical system that produces this distribution.\cite{remark0}
Finally, we remark that we introduced the shift parameter $s$ for the generating function of factorial cumulants since we aim at identifying deviations from uncorrelated statistics.
In principle, however, such a shift parameter can also be introduced for the generating function of ordinary cumulants.\cite{remark2}

% - - - - - - - - - - - - - - - - - - - - - - - - - - - - - - - - - - - - - -
\section{Cumulants in the short-time limit}\label{sec:factorialstatics}

From now on, we concentrate on charge-transfer statistics in Coulomb-blockade systems in the short-time limit.
The latter is defined by the length $t$ of the measuring time interval being small as compared to the average waiting time between two adjacent tunneling events.
Of course, $t$ has to remain larger than the time resolution of the detector.
We assume that the duration of all the fundamental tunneling processes (sequential or Andreev tunneling) is shorter than the time resolution of the detector.
In this case, the system's dynamics can be described by a Markovian master equation.
Therefore, transport regimes in which non-Markovian effects become relevant\cite{braggio_full_2006, flindt_counting_2008, flindt_counting_2010} are not covered by the following analysis.
Furthermore, we require that the charge-transfer statistics is measured after the system has reached its stationary state.
Charging energy limits the number of possible charge states.
However, we allow for arbitrary many microscopic realizations of each charge state (e.g., the electrons on the island may occupy different orbital and/or spin states).

We aim at identifying correlations between the transfers of individual electrons.
In our example, there are two sources of correlations.
First, correlations between two sequential tunneling events can arise as a consequence of the charging energy since the probability for a given transition depends on the charge state of the island.
Second, for Andreev tunneling there is, due to superconducting pairing interaction, a correlation between the two electrons tunneling simultaneously within one fundamental charge-transfer process.

{\it Ordinary} cumulants in the short-time limit are {\it not} a convenient tool to identify these correlations. 
This can be understood in the following way.
Let us assume that a single electron can tunnel with a rate~$\Gm_1$, two electrons (Andreev tunneling) with $\Gm_2$, and, in generalization, there may be fundamental processes in which even a larger number $N$ of electrons tunnel simultaneously with $\Gm_N$.\cite{remark4}
In the short-time limit, the full counting statistics is given by $P_N(t)= \Gm_N  t + {\mathcal O}(t^2)$ and $P_0(t)=1-\sum_{N\geq 1} \Gm_N t + {\mathcal O}(t^2)$. 
It contains only contributions with at most {\it one} fundamental $N$-particle process during the measuring-time interval; contributions with more such processes are of order $t^2$ or higher.
Expanding the logarithm of the generating function up to linear order in time, $\ln{\cal M}(z,t) = \sum_{N\geq 1} (e^{N z} -1)\Gm_N t + {\mathcal O}(t^2)$, we obtain the short-time limit of the ordinary cumulants,
\begin{equation}
	C_m(t) = \sum_{N\geq 1} N^m \Gm_N t \, .
\end{equation}
They are all positive and linear in $t$, irrespectively of the presence or absence of correlations.
Only the relative magnitude of the cumulants of different order $m$ may be used to identify the presence of fundamental tunneling processes with $N\ge 2$.

This is qualitatively different for the short-time behavior of factorial cumulants.
For a Poisson binomial distribution, the probabilities $p_j(t)$ are positive and depend, in the short-time limit, linearly on $t$.
Together with Eq.~(\ref{eq:cumwithpj}), this fixes both the sign and the time dependence, $C_{{\rm F}, m}(t)\propto (-1)^{m-1}t^m$.
Both a different sign and/or a different time dependence must be due to correlations.
In the following, we illustrate this for Coulomb-blockade systems with different numbers of accessible charge states and different fundamental tunneling processes.

As $t$ goes to $0$, no charge is transferred, $P_N(0)=\delta_{N,0}$, and all (generalized) factorial cumulants vanish.
For small but finite $t$, the $P_N(t)$'s in Eq.~(\ref{eq:generating_factorial}) with $N\ge 1$ are small, which implies that the zeros $z_j$ of the generating function have a large magnitude.
From Eq.~(\ref{eq:cumulantscos}) it is clear that the short-time behavior of the (generalized) factorial cumulants is governed by those zeros $z_j(t)$ that diverge slowest as $t\rightarrow 0$.
To determine them, we expand the generating function, Eq.~(\ref{eq:generating_factorial}), in powers of $t$ and keep only the leading terms proportional $\propto t^0$. Hence, the constant term $z^0 t^0$ must be taken into account plus those terms $\propto z^kt^l$ with the smallest appearing ratio $l/k$ (for $l\ge 1$).
This yields $z_j(t)\propto t^{-l/k}$.

% - - - - - - - - - - - - - - - - - - - - - - - - - - - - - - - - - - - - - -
\subsection{Sequential tunneling for two charge states}\label{sec:shortsq}

We start by discussing sequential tunneling between two accessible charge states.
In the short-time limit, we can approximate the generating function by $M_{\rm F}(t)\approx 1+z P_1(t)$ where $P_1(t)=\Gm_1 t$ linear in time describes a single tunneling-out event.
Contributions $P_N(t)$ with $N\ge 2$ tunneling-out events can be ignored for the following reason.
After each tunneling-out event (being counted) an electron has to tunnel in (which is not counted) before another tunneling-out event can happen. 
This implies $P_{N}(t)\propto t^{2N-1}$ (with $N\ge 1$), and from all the contributions $z^N P_{N}(t)\propto z^N t^{2N-1}$ to the generating function, only the one with $N=1$ has to be kept since this has the smallest ratio $2-1/N$ of the exponents for $t$ and $z$.
Thus, there is only one zero $z_1(t)=-1/(\Gm_1 t)$ and by means of Eq.~(\ref{eq:cumulantscos}) we get the short-time generalized factorial cumulant 
\begin{equation}
\label{eq:cumseq2}
	C_{s, m}(t)=(-1)^{m-1} (m-1)! \,  \left( \Gm_1 t\right)^m \propto t^m\, ,
\end{equation}
independent of $s$.
Despite the presence of interactions via the charging energy, we find that both the sign $(-1)^{m-1}$ and the power law $t^m$ in the short-time limit are the same as for a Poisson binomial distribution of independent tunneling events.
This seems to be a consequence of the restriction to two charge states only.

The short-time behavior of recently measured factorial cumulants for hole transfer in semiconductor quantum dots with dense excitation spectrum\cite{komijani_counting_2013} is in full agreement with Eq.~(\ref{eq:cumseq2}): while the alternation of the sign with increasing $m$ is explicitly commented on, also the power-law dependence~$t^m$ is clearly visible in the data shown in Fig.~5(e) of Ref.~\onlinecite{komijani_counting_2013}.
Our paper now provides an explanation for this experimental finding.

\subsection{Sequential tunneling for three charge states}\label{sec:shortsq3}

If three charge states are accessible, two sequential-tunneling processes transferring an electron from the island into the leads can occur in a row, i.e., $P_2(t) \propto t^2$ and $z^2 P_2(t)\propto z^2 t^2$ is equally important for the position of the zeros of the generating function as $z P_1(t)\propto z t$. 
But still, terms with $P_{N\ge 3}(t)\propto t^{2N-2}$ are negligible since they involve at least $N-2$ tunneling-in events that are not counted. 
As a result, we get $M_{\rm F}(t)\approx1+z P_1(t)+z^2 P_2(t)$ with two zeros $z_{1,2}(t)=-2/\left [ P_1(t) \pm \sqrt{P_1^2(t)-4P_2(t)}\right]$,
which yields 
\begin{equation}\label{eq:cumseq3}
	C_{s, m}(t) =(-1)^{m-1} (m-1)!\, \left[ P_1(t)\right]^m a_m \propto t^m \, ,
\end{equation}
independent of $s$, where
\begin{equation}
	a_m =  \sum_{j=\pm1} \left( \frac{1}{2}+\frac{j}{2}  \sqrt{1-\frac{4P_2(t)}{P_1^2(t)}}\, \right)^m 
\end{equation}
is independent of time.
Again, in the short-time limit, $C_{s, m}(t)$ obeys the power law $t^m$.
But now, the sign can be modified: for $4P_2(t)> P^2_1(t)$ there are some $m$'s for which $a_m<0$ such that $C_{s,m}(t)$ has the opposite sign as for a Poisson binomial distribution of independent tunneling events.
We conclude that for a detection of correlations in the full counting statistics of sequential tunneling, more than two charge states are necessary.

It is straightforward to extend this procedure to include more charge states.
For each extra charge state, one more term $z^N P_N(t)\propto z^N t^N$ needs to be kept in the generating function, Eq.~(\ref{eq:generating_factorial}), giving rise to one more zero $z_j(t)\propto 1/t$ such that $C_{s,m}(t)\propto t^m$ independent of $s$ but with a sign that may or may not be given by $(-1)^{m-1}$.

% - - - - - - - - - - - - - - - - - - - - - - - - - - - - - - - - - - - - - -
\subsection{Sequential plus two-electron tunneling for three charge states}
\label{seq:shortgeneral}

We now show that the presence of two-electron processes such as Andreev tunneling, where two electrons are tunneling out of the island simultaneously, leads to a power-law behavior of the short-time (generalized) factorial cumulants different from $t^m$.
While $P_1(t)=\Gm_1 t$ is given by sequential tunneling with rate $\Gm_1$, $P_2(t)=\Gm_2 t$ is dominated by two-electron tunneling with rate $\Gm_2$.
Sequential tunneling of two electrons is negligible since it is of order $t^2$.
Keeping only those terms $\propto z^k t^l$ with the smallest ratio $l/k$, we approximate the generating function for the case of three charge states by $M_{\rm F}(t)\approx1+z^2 \Gm_2 t$.
Plugging the zeros $z_{1,2}(t)=\pm i /\sqrt{\Gm_2 t}$ into Eq.~(\ref{eq:cumulantscos}) yields, for even $m$, the short-time (generalized) factorial cumulants
\begin{equation}
\label{eq:cumcoteven}
	C^{\rm{even}}_{s,m}(t)= - (m-1)! \, 2\left( -\Gm_2 t\right)^{\frac{m}{2}} \propto t^{\frac{m}{2}} \, ,
\end{equation}
independent of $s$ and independent of $P_1(t)$.
For odd $m$, however, the two zeros determined above cancel out each other when plugged into Eq.~(\ref{eq:cumulantscos}), indicating that, for odd $m$, the approximation of the generating function was too crude.
Instead, we need to include the next-order correction.
To identify which terms to include, we use that the zeros diverge with $z_j(t)\propto 1/\sqrt{t}$.
Thus, the correction (of order $\sqrt{t}$) is introduced by terms $\propto zt$.
Therefore, we get $M_{\rm F}(t)\approx1+z \Gm_1 t +(z^2+2z) \Gm_2 t$, while all other terms can still be neglected. 
The corresponding zeros are $z_{1,2}=\pm i / \sqrt{\Gm_2 t} - \Gm_1/(2\Gm_2)-1$, which yields $1/[z_{1,2}(t)-s+1]= \mp i \sqrt{\Gm_2 t}-(\Gm_1/2 + s \Gm_2) t$, and, finally, 
\begin{equation}
\label{eq:cumcotodd}
	C^{\rm{odd}}_{s,m}(t)= m! \,\left( -\Gm_2 t \right)^{\frac{m-1}{2}}\left(\Gm_1 +2 s \Gm_2 \right) t \propto t^{\frac{m+1}{2}} \, .
\end{equation}
We note that, in contrast to even $m$, the odd-$m$ generalized factorial cumulants depend both on $s$ and on the sequential-tunneling rate $\Gm_1$.

Combining the results for even and odd $m$, we conclude that the time-dependence $C_{s,m}(t) \propto t^{\left \lceil m/2\right \rceil}$, where $\left \lceil m/2\right \rceil$ is the smallest integer larger or equal to $m/2$, is qualitatively different from a Poisson binomial distribution of independent tunneling and also from a Coulomb-blockade system in which only sequential tunneling occurs.
Also the sign is different, for $s\gtrless -\Gm_1/(2\Gm_2)$ given by $(-1)^{\left \lceil \pm m/2\right \rceil-1}$.
We remark that the relative strength of single- and two-electron tunneling, $\Gm_2/\Gm_1$, influences the time below which the short-time limit is applicable but, ultimately, two-electron tunneling will always dominate the (generalized) factorial cumulants.

% - - - - - - - - - - - - - - - - - - - - - - - - - - - - - - - - - - - - - -
\subsection{Higher-order tunneling}

The discussion can be easily extended to include tunneling processes in which up to $N_\tx{max}$ electrons simultaneously leave the island. 
We find that the tunneling processes with $N_\tx{max}$, i.e., the largest number of simultaneously transferred electrons, dominate the short-time behavior of the generalized factorial cumulants.
The time dependence is given by $C_{s,m}(t) \propto t^{\left \lceil m/N_\tx{max}\right \rceil}$.
For $m=N_\tx{max},2N_\tx{max},3N_\tx{max},\ldots$, the generalized factorial cumulants depend only on the leading order of the zeros $z_j(t) \propto t^{-1/N_\tx{max}}$, and they are independent of $s$.
The zeros and, thus, the $p_j(t)=-1/z_j(t)$ are complex, and an interpretation of the charge-transfer statistics in terms of a Poisson binomial distribution of independent tunneling events is impossible.

\begin{figure}[b]
\includegraphics[scale=1.10]{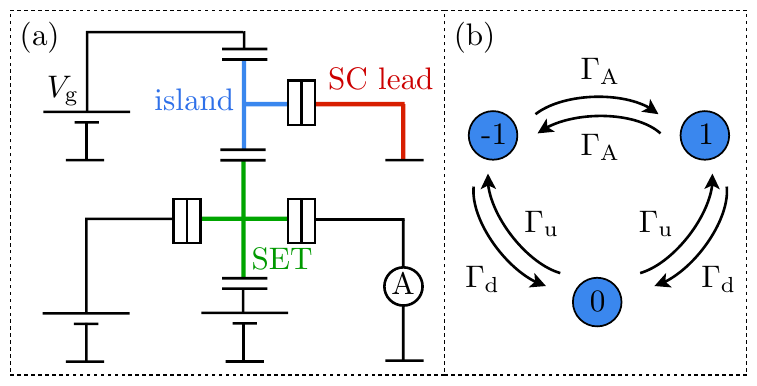}
\caption{
	(Color online) (a) Normal-state metallic island (blue) weakly tunnel coupled to a superconducting lead (red). Via the gate voltage $V_\tx{g}$, the equilibrium charge $n_\tx{g}$ of the island is tuned to zero. The current through a single-electron transistor (green) monitors the island charge $n$ as a function of time. (b) Sketch of the states and transition rates. 
		}
\label{fig1}
\end{figure}

% - - - - - - - - - - - - - - - - - - - - - - - - - - - - - - - - - - - - - -
\section{Sequential and Andreev tunneling in a single-electron box}\label{sec:model}
\begin{figure*}
{\includegraphics[scale=1.0]{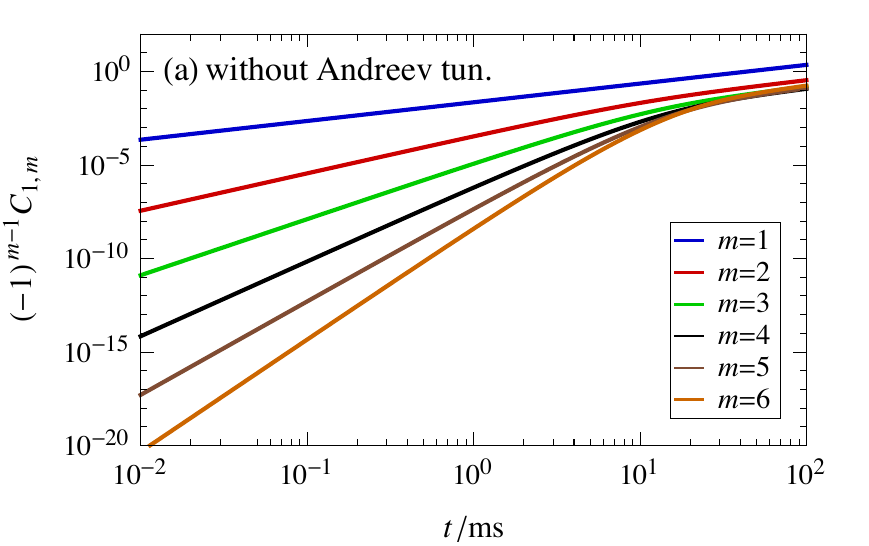}}
{\includegraphics[scale=1.0]{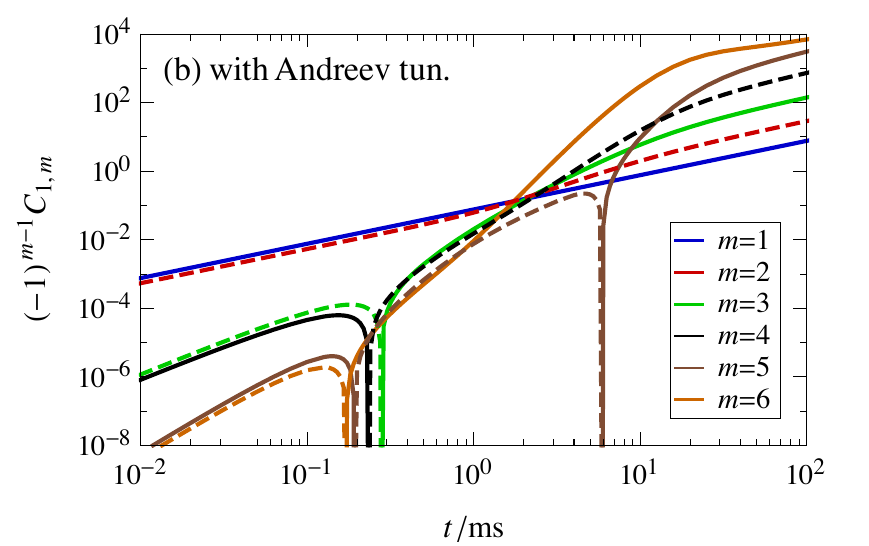}}
\caption{
	(Color online) Factorial cumulants as a function of time (a) without and (b) with Andreev tunneling. The sign of $(-1)^{m-1}C_{1,m}$ is positive for continuous and negative for dashed lines.
	}
\label{fig2}
\end{figure*}
We illustrate our findings for a model system that has been experimentally realized in Refs.~\onlinecite{saira_environmentally_2010, maisi_real_2011, maisi_full_2014}. 
Its setup is shown in Fig.~\ref{fig1}(a). 
A normal-metal island is tunnel coupled to a superconducting lead. This single-electron box (SEB) is characterized by the (normal-state) tunnel resistance $R_\tx{T}$ and the charging energy $E_\tx{C}(n-n_\tx{g})^2$
for $n$ electrons on the island. 
Via a gate voltage $V_\tx{g}$, the continuously variable gate charge $n_\tx{g}$ is tuned to the symmetry point $n_\tx{g}=0$.
To monitor the integer charge $n$ on the island, a single-electron transistor (SET) is electrostatically coupled: for each value of $n$ there is a characteristic value of the current through the SET.
At low temperature, only three charge states play a role, $n=-1,0,1$ (relative to some background).
Single-electron tunneling leads to transitions between $n=0$ and $\pm 1$, and Andreev tunneling imply direct changes between $n=+1$ and $-1$.
The corresponding rates are $\Gm_\tx{u}$, $\Gm_\tx{d}$, and $\Gm_\tx{A}$, see Fig.~\ref{fig1}(b).

In Ref.~\onlinecite{maisi_full_2014}, the full counting statistics of Andreev tunneling (without distinguishing tunneling in from tunneling out and without counting sequential tunneling events) was measured. 
A strongly super-Possonian distribution was found and attributed to avalanches of Andreev processes that form due to the interplay of Andreev and single-electron tunneling.
By interpreting the data in terms of the cumulant generating function in the long-time limit, avalanches of up to 20 Andreev processes have been identified.

In the following, we consider the very same system but study the {\it short-time} charge-transfer statistics for {\it tunneling out} of the island via {\it both sequential and Andreev tunneling} and characterize it via {\it generalized factorial cumulants}.
Instead of calculating the tunneling rates $\Gm_\tx{u}$, $\Gm_\tx{d}$, and $\Gm_\tx{A}$ in the presence of an electromagnetic environment,\cite{saira_environmentally_2010, pekola_environment_2010} we take the values $\Gm_\tx{u}=12\ \tx{Hz}$, $\Gm_\tx{d}=252\ \tx{Hz}$, and $\Gm_\tx{A}=615\ \tx{Hz}$ that were experimentally measured in Ref.~\onlinecite{maisi_full_2014} at temperature $50\, \tx{mK}$ for $E_\tx{C}=40 \mu \tx{eV}$, $\Delta=210 \mu \tx{eV}$, and $R_\tx{T}=490\  \tx{k} \Omega$.

% - - - - - - - - - - - - - - - - - - - - - - - - - - - - - - - - - - - - - -
\subsection{Master-equation approach}\label{sec:technique}
The charge-transfer dynamics can be described by the Markovian master equation 
\begin{equation}
\begin{aligned}
	\dot{P}_N^{-1}(t)&=-\left(\Gm_{\tx A} + \Gm_{\tx d}\right)P_N^{-1}(t) + \Gm_{\tx u}P_{N-1}^0(t) + \Gm_{\tx A}P_{N-2}^1(t)\, ,\\
	\dot{P}_N^0(t)&= \Gm_{\tx d}P_N^{-1}(t) -2\Gm_{\tx u}P_N^0(t)+\Gm_{\tx d} P_{N-1}^1(t)\, ,\\
	\dot{P}_N^1(t)&=\Gm_{\tx A}P_N^{-1}(t) + \Gm_{\tx u}P_{N}^0(t)  -\left(\Gm_{\tx A} + \Gm_{\tx d}\right)P_{N}^1(t),
\end{aligned}
\end{equation}
for the probability $P_N^n(t)$ that at time $t$ the island is in charge state $n$ and $N$ electrons have left the island in time interval $[0,t]$.
The counting of $N$ is done by identifying the transitions between different charge states $n$ in the time trace of the measured SET current.

By performing the $z$ transform, i.e., multiplication with $z^N$ and then summation over $N$, we can rewrite the master equation in matrix-vector notation
\begin{equation}\label{eq:masterZ}
	\dot{\vec{P}}_{z}(t)= \vec{W}_{z} \vec{P}_{z}(t) \, ,
\end{equation}
with vector $\vec{P}_{z}=\sum_{N=0}^\infty z^N ( {P}_{N}^{-1},{P}_{N}^{0},{P}_{N}^{+1})^T$ and matrix
\begin{equation}
	\vec{W}_{z} =\begin{pmatrix} -\Gm_{\tx A} - \Gm_{\tx d}& z \Gm_{\tx u} & z^2 \Gm_{\tx A}\\ \Gm_{\tx d}& -2\Gm_{\tx u} & z \Gm_{\tx d}\\ 	\Gm_{\tx A}&\Gm_{\tx u} & -\Gm_{\tx A} -\Gm_{\tx d} \end{pmatrix} \, .
\end{equation}
The solution of Eq.~\eqref{eq:masterZ} is $\vec{P}_{z}(t) = \exp \left( \vec{W}_{z}t \right) \vec{P}_z(0)$.
Since $P_N^n(0)\sim \delta_{N,0}$, the initial vector $\vec{P}_z(0)$ is independent of $z$ and describes the initial probability distribution.
Assuming that the system has reached its steady state before electron counting starts, $\vec{P}_z(0)$ is determined by $\vec{W}_1 \vec{P}_z(0)=0$ and $\vec{e}^T\cdot \vec{P}_z(0)=1$, with $\vec{e}^T=(1,1,1)$. 
Generalized factorial cumulants are given by $C_{s, m}(t)=\partial_z^m [\ln {\cal M}_{s}(z,t)] |_{z=0}$ with ${\cal M}_s(z,t)= [\vec{e}^T \cdot \vec{P}_{z+s}(t)]/[\vec{e}^T \cdot \vec{P}_{s}(t)]$.

% - - - - - - - - - - - - - - - - - - - - - - - - - - - - - - - - - - - - - -
\subsection{Factorial cumulants without Andreev tunneling}
\label{sec:modelshort}

In order to identify the influence of Andreev tunneling on the full counting statistics, we, first, calculate factorial cumulants ($s=1$) in the absence of Andreev tunneling (we set $\Gm_\tx{A}=0$ but keep $\Gm_\tx{u}$ and $\Gm_\tx{d}$ unchanged). 
In Fig.~\ref{fig2}(a), we plot $(-1)^{m-1}C_{1,m}(t)$ as a function of the length $t$ of the measuring interval. 
Due to the prefactor $(-1)^{m-1}$, the result is positive for all $m$ and $t$, indicating that in the studied scenario the full counting statistics can be described by a Poisson binomial distribution of independent tunneling events.
In the long-time limit, all factorial cumulants are linear in time because $C_{1,m}(t) \to  \partial^m_z[ \lambda (z)]_{z=1} t$, where $\lambda$ is the eigenvalue of $\vec{W}_z$ with the largest real part at $z=1$.

In the short-time limit (given by $\Gm_\tx{d} t \ll 1$), however, we find $C_{1,m}(t) \propto t^m$, as derived in Sec.~\ref{sec:factorialstatics}.
In fact, the numerical results are very well described by Eq.~\eqref{eq:cumseq3} with
\begin{equation}
	P_1(t)= \frac{2\Gm_\tx{u}\Gm_\tx{d}t}{ 2\Gm_\tx{u}+\Gm_\tx{d}}
	\, , \quad 
	P_2(t)= \frac{\Gm_\tx{u}^2\Gm_\tx{d}t^2}{2( 2\Gm_\tx{u}+\Gm_\tx{d})}\,.
\end{equation}
We remark that the short-time (but not the long-time) factorial cumulants in the absence of Andreev tunneling are identical to those when Andreev tunneling processes are present but not counted.
This could be immediately realized for the set up of Ref.~\onlinecite{maisi_full_2014}.
Furthermore, we mention that if we chose for the considered model $\Gm_\tx{u}>\Gm_\tx{d}/2$ then $4P_2(t)>P_1^2(t)$, and $(-1)^{m-1}C_{1,m}(t)$ would become negative for some $m$ and $t$, in accordance to the discussion in Sec.~\ref{sec:shortsq3}.
\begin{figure*}
{\includegraphics[scale=1.0]{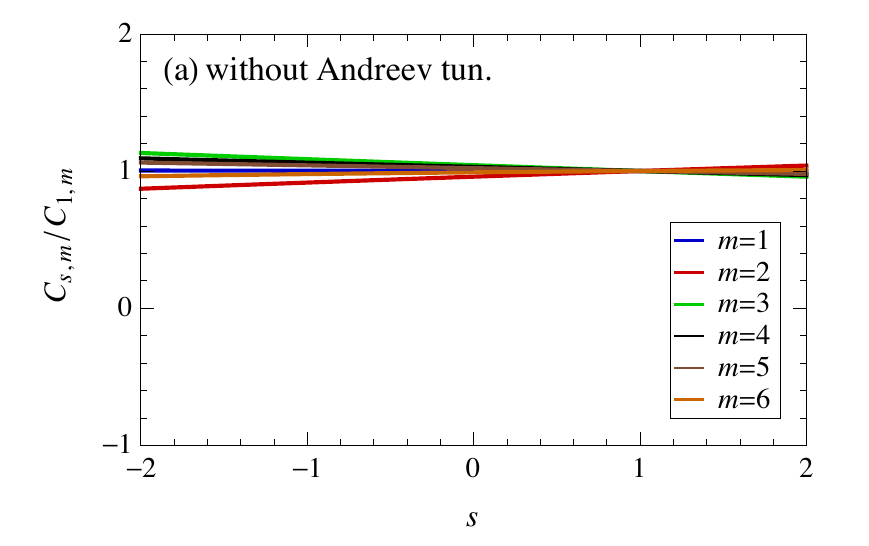}}
{\includegraphics[scale=1.0]{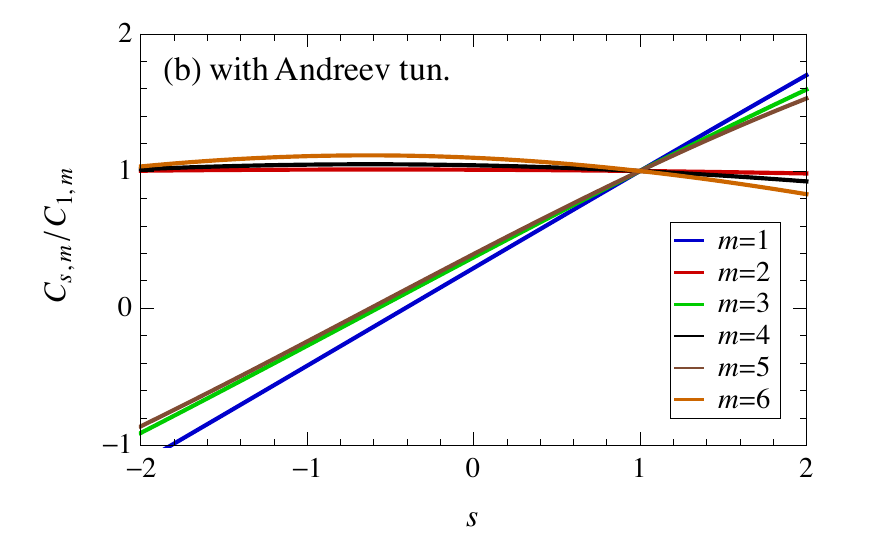}}
\caption{
	(Color online) Generalized factorial cumulants $C_{s, m}$ normalized by $C_{1,m}$ as a function of $s$ (a) without and (b) with Andreev tunneling at time (a) $t=0.7 \, \tx{ms}$ and (b) $t=0.1 \, \tx{ms}$. In (b), factorial cumulants of odd order $m$ display an $s$~dependence.}
\label{fig3}
\end{figure*}

\subsection{Factorial cumulants with Andreev tunneling}

In the presence of Andreev tunneling, $(-1)^{m-1}C_{1,m}(t)$ looks qualitatively different, see Fig.~\ref{fig2}(b).
The only common feature is the linear long-time behavior.
We now find extended regions with negative $(-1)^{m-1}C_{1,m}(t)$, indicated by dashed lines in Fig.~\ref{fig2}(b).
Furthermore, in the short-time limit, the numerical results are no longer described by Eq.~\eqref{eq:cumseq3} but by Eqs.~\eqref{eq:cumcoteven} and \eqref{eq:cumcotodd} with
\begin{equation}
	\Gm_1 = \frac{2\Gm_\tx{u}\Gm_\tx{d}}{ 2\Gm_\tx{u}+\Gm_\tx{d}}
	\, , \quad 
	\Gm_2 = \frac{\Gm_\tx{u} \Gm_\tx{A}}{2\Gm_\tx{u}+\Gm_\tx{d}}\, ,
\end{equation}
i.e., we get the power law $C_{1,m}(t) \propto t^{\left \lceil m/2\right \rceil}$.

\subsection{Generalized factorial cumulants}

Finally, we investigate the $s$ dependence of the generalized factorial cumulants. 
In the absence of Andreev tunneling, the $s$ dependence of the generalized factorial cumulants should vanish in the short-time limit, see Eq.~(\ref{eq:cumseq3}).
This is in agreement with the numerical results for $C_{s, m}/ C_{1,m}$ as a function of $s$, shown in Fig.~\ref{fig3}(a):
the displayed $s$ dependence is very weak, which indicates that the chosen $t$ is already well in the short-time regime.

In the presence of Andreev tunneling, the generalized factorial cumulants of even order $m$ should be $s$ independent while for odd order $m$ a linear $s$ dependence is expected, see Eqs.~(\ref{eq:cumcoteven}) and (\ref{eq:cumcotodd}). Also this agrees with the numerical simulations, see Fig.~\ref{fig3}(b).

We conclude by commenting on the extra information that the $s$ parameter in the generalized factorial cumulants provides.
As discussed in the derivation of Eq.~(\ref{eq:cumcotodd}), the $s$ dependence of some short-time cumulants stems from the next-to-leading order of the time dependence of the zeros $z_j(t)$ of the generating function since the leading order does not contribute to these cumulants.
This observation is not limited to the specific example discussed here, but holds true in general, as can be directly seen from Eq.~(\ref{eq:cumulantscos}).

\section{Conclusions}

In this paper, we address the question of how to identify correlations among electron tunneling events in the time-dependent charge-transfer distribution $P_N(t)$.
A measured deviation from a Poisson distribution only excludes the possibility of independent tunneling events with {\it identical} tunneling rates.
To prove correlations, one must also exclude Poisson binomial distributions of independent but {\it nonidentical} tunneling events.
This is difficult to do when analyzing the charge-transfer statistics in terms of ordinary cumulants commonly used in literature.
Factorial and generalized factorial cumulants, on the other hand, are quite suited to perform this task, since both their sign and their short-time power-law dependence turn out to be good indicators of correlations.
In the absence of correlations, the sign of the factorial cumulants alternate with order, $(-1)^{m-1} C_{{\rm F}, m}(t) > 0$, and in the short-time limit the time dependence displays a $t^m$ behavior.
The measurement of a different sign or a different short-time behavior for any factorial cumulant would be incompatible with uncorrelated electron transport. 

Correlations may be induced by the charging energy in Coulomb-blockade systems,~\cite{stegmann_detection_2015} but also by attaching superconducting leads.
The presence of Andreev tunneling changes the time dependence of the factorial cumulants drastically: instead of $t^m$, we predict a power law $t^\frac{m}{2}$ for even and $t^\frac{m+1}{2}$ for odd orders $m$ in the short-time limit.

We conclude by commenting on the experimental feasibility of our proposal.
Determining factorial instead of ordinary cumulants does not introduce any extra complication or the need for higher accuracy.
Furthermore, cutting the total measured time trace into shorter and, therefore, more time intervals to achieve the short-time limit even reduces statistical errors.~\cite{remark1}

% - - - - - - - - - - - - - - - - - - - - - - - - - - - - - - - - - - - - - -
\acknowledgments
We thank Alfred Hucht and Stephan Wei{\ss} for helpful discussions and acknowledge financial support from the DFG under project KO 1987/5.


\begin{thebibliography}{50}
\bibitem{remark0}
We remark that for any experimental analysis of full counting statistics, the convergence of the series Eq.~(\ref{eq:generating_factorial}) and the product Eq.~(\ref{eq:prodzj}) is not an issue since $N$ can be considered to be limited by some large but finite cutoff $N_\text{cutoff}$.
Then, Eq.~(\ref{eq:generating_factorial}) is a finite polynomial with a finite number of zeros, leading to a finite product Eq.~(\ref{eq:prodzj}).
In this respect, Eq.~(\ref{eq:prodzj}) is general.
Theoretically, there are distributions for which the factorization fails. 
An example is the Poisson distribution which does not possess any zero.
But, in any experiment with a finite $N_\text{cutoff}$, the Poisson distribution can only be obtained approximatively as a Poisson binomial distribution for which the factorization holds.
\bibitem{remark4}
In general, there may be several $N$-particle processes whose individual tunneling rates $\Gamma_{N\chi}$ depend on the state $\chi$ of the island. 
In this case, $\Gamma_N$ is the average rate, where each $\Gamma_{N\chi}$ is weighted with the probability to find the island in state $\chi$.
\end{thebibliography}
\end{document}